# THE SOUTH AFRICAN SOFTWARE INDUSTRY AS A KEY COMPONENT OF ECONOMIC DEVELOPMENT: PIPEDREAM OR POSSIBILITY?

*Patrick Mukala*
*Software Engineering*
*Tshwane University of Technology*
*October, 2009*

**Abstract**:
The Information and Communication sector has undoubtedly played a pivotal role in changing the way people live nowadays. Almost every area of our lives is affected by the presence and the use of the new information and communication technologies. In this regard, many researchers' attention has been attracted by the influence or the significant impact of these technologies on economic growth and development. Although the history of South Africa has had some drawbacks that could constitute a big obstacle to the emergence of a successful economic environment, the actual status of the country regarding its economy and the role that it plays in Africa towards the rest of the African countries is a vital example of an emerging economic force in Africa. This paper examines the crucial role that ICT has played and is still playing in the South African economy growth and more specifically the significance of the economic effects of the software industry. It makes use of the framework used by Heavin et al. (2003) to investigate the Irish software industry in order to analyze the impact of endogenous factors-national, enterprise and individual - on the software industry and its implication on the economic growth in South Africa.

**Keywords**: South African Software Industry, economic growth, ICT sector, government intervention, exogenous (external) factors, endogenous (internal) factors.

## I.   Introduction

The survival of any country in times in crisis and its potential aim to its global development are critical factors that constitute key roles of any government in any nation world wide. The economic area of the country constitutes its enormous potential for its development and well being of its population. On the other hand, supporting the economy requires a deep and careful management of resources at the country's disposal as well as the contribution of other vital areas of the country's asset such us industry. The case of South Africa's economy constitutes the main focal point of its paper; the importance and diverse industrial means that are in place to sustain this country's economic force for the well being of its 44 millions population forms a complete interest of many researchers.

The information and Technology sector and telecommunications are nowadays regarded as a potential cornerstone to economic boost. The software industry in particular has had an increasing impact on the development and spectacular growth of economic sectors in many countries world wide. Considered as one of major industries in the world, the software industry has successfully contributed to the development and significant growth of the economic sectors in an impressive growing number of countries such as Singapore, China, India and most definitely Ireland. The latter's changes in its economic development is an inspiring example from which this paper draws a perspective view on how concentrating on Information Technology and especially the software industry in the economic policies of the government has a significant impact on the growth of the economic area towards the country's industrial development.

In their analysis of the successful influence of the software industry on the Irish economic growth, Heavin et al. (2003) support that any other countries that would like to strategically adopt software industry as a key component towards accomplishing economic development should understand the interaction between both policy and socio-cultural factors. This just implies that these countries that aim to achieve the same results or grow their economic force need to have a better comprehension of the vital factors that have critically influenced the emergence of Ireland's software industry and thus, these findings can provide significant lessons in order to achieve the same goal.

The software industry, as a global industry, is inextricably link with hardware technologies; that is towards its implication on the economic development, this impact is also felt on the contribution of the manufacturing industry, especially hardware manufacturing companies. Hardware applications are developed to suit the needs of emerging software packages. There is an interchangeable influence of both software applications and hardware designs. The implications of software industry are very significant on almost every area of the Information and Communication technology (ICT) sector. Thus, the contributions of the entire ICT on the development and growth of the economy clearly demonstrates indirectly the impact of the software industry as the latter plays a significant role when we refer to the new information and communication technologies in general. One cannot talk about the Information and Communication Technologies without referencing the software industry.

How policies regarding economic development are implemented, is undeniably a key facilitator for an effective economic takeoff and successful growth. As mentioned previously, the Information and Communication Technology area should be critically considered and highly prioritized in implementing these policies. Let have a look at how these are perceived in South Africa.

The Information Technology field, as sustained by a great number of published research reports has constantly attracted the attention of the South Africa's government regarding its approach its potential contribution to economic growth and development (Breitenbach et al. , 2005).

In promoting and sustaining the implementation of the Information and Communication Technology in South Africa, a range of initiatives has been considered to excite its growth and thus contribute to the economic development. One of these initiatives is the SAITIS project (South African Information Technology Industry Strategy) project.

According to a report published by the department of Trade and Industry in 2000 on "South African ICT sector development framework Summary", the South African Information Technology Industry Strategy (SAITIS) project was developed in 1994. In this report, it is sustained that new processes needed to be put in place during that transition period to replace the Apartheid structures in order to empower South Africa. The main objective of the project was "to bridge the global and local development gaps and develop a robust, growing and sustainable South African Information and Communications Technology Sector that would directly support and contribute to the challenges of sustainable economic growth, social upliftment and empowerment"(Department of Trade and Industry,2000).

According to SAITIS, the ICT area of South Africa includes major industries such as Hardware and Telecommunications equipment manufacturing companies, and a wide range of services that include services that deal with software applications development, software packages products and services that directly concern the telecommunication area. All these areas have increasingly contributed to the country's economic revenue. An example in this regard would be the case of telecommunication infrastructures in 2002; their contribution was estimated to R74 billion compared to only R7 billion in 1992.

Former President Thabo Mbeki while addressing the people in the official State of Nation Address in 2007, pointed out a number of precise goals and objectives that need to be reached in order to boost South Africa to a better standard level, and to fulfill the vision of an improved and excellent life for all South Africans as a result of a strong economic country's status. These specific set goals include among other factors how to raise the rate of investment especially in the first economy, enhance the possibility of doing business by reducing cost and promote the development of the small and medium business sector. All these goals can only be possible if technology is considered as a pivotal factor and key component of the economic growth.

In the following lines, our main focus will be to demonstrate the direct and indirect implications that the ICT sector, more specifically the software industry has had on the South African economic force development and future possible trends that this industry has on the entire economic sector of the country. We look at different possible internal or external factors that contribute to the emergence of this economic force; what approaches have been considered by the economic policy planners or the South African government to sustain the development of a strong software industry with a long-term vision of the implications of the latter on strengthening the country's economic force.

## II. Literature Review

Information technology (IT) has been and continues to draw attention from researchers and economical analysts on how it affects economic growth in developing countries, especially Africa. As previously mentioned, in the midst of the collection of technologies that make up the industry software is very important since the rest of technologies do critically require software so as to function, and most of them are even derived from software. The software industry forms without any doubt an essential if not the most essential component of the overall value brought by the information and communication technologies as a whole.

There is a view that prevails in economics and this view supports that economic growth is without a doubt driven by advances and innovation that take place in technology because this enables a more efficient production. The direct effect of technology revolutions can be perceived through Information and Communication Technology; hence, for this ultimate reason they are considered as an essential and critical factor to driving economic development nowadays (Pohjola, 2002).

Dr. Nagy Hanna of the World Bank and his partners in their research on national software industry development strongly believe that Information and communication technologies and related services have and still are contributing to innovative changes in business and citizens' everyday life. If this does not happen in developing countries, meaning if these technologies are not significant in developing countries, a big gap will be created between these people and the global information society. These technologies enormously impact on social and economic development (Hanna et al., 2003).

From a traditional point of view, researchers have explained the economic success of any national industry based on a couple of terms from macro-economy. These related terms include among others interest and exchange rates, the value of natural resources, government's policy and practical implications in the market force…The last factor namely the government intervention is the cornerstone to promoting a strong economic politics through sustaining software industry.

For almost a decade, software production has been considered as an exclusive industry and its implication perceived nowadays as an important and essential for the successful development of developing countries' economies and of nowadays an industry, essential for the growth of the economies of the developing countries; and the initiation of plans to promote software industries in these countries is a main concern (Heeks, 1996).

Because of their success in implementing software industry as a critical element of economic initiatives empowerment, a World Bank report has classified a number of countries based on a set of four factors with regards to software production: China for its strategy regarding the cost, Singapore and Ireland with regards to the language used as they are English-speaking nations, India, Ireland, Singapore and Israel for promoting a favorable atmosphere that enables doing business (Carmel, 2003).

In this regard, the government's action on supporting the growth of software industry has a great significance on the impact of the latter on the country's economic development; its practical contribution would be through supporting higher education, providing funds and promoting research activities in accordance with technology and economic growth, and most importantly endowing regulations that are favorable to investment in corporate (Shirley et al., 1997).

In recent researches' reports, it is sustainable that technology and innovative activities enable an enhanced economic growth in developing countries and these reports indicate that there exists a major rapprochement between both growth in productivity and progress in technology (Breitenbach et al., 2005).

From a study on how information technology impacts economic development in developing countries, Professor Chrisanthi Avgerou argues that most of modern theories on economic change do actually recognize the importance of information and communication technologies (Avgerou, 1998). In order to get the meaning of economic development and technical innovation, he points out the Noe-Schumpeterian theory.

This cited theory explains how certain technologies have a real persistent impact in economic development. Thus, it is implied in this theory that "*a given technology have impact on economy if it engenders a number of new products and services; it reduces the actual costs and contributes to the improvement of used processes, services and products of many sectors of the economy; gains widespread social acceptance; and generates strong industrial interest as means for profitability and competitive advantage*" (Avgerou, 1998).

From this perspective, when considering the Information Technology domain with the software industry as its centre as mentioned previously in this paper, we get the idea that numerous studies have been conducted to understand the link between development with regards to economic growth and technology more specifically information and communication technologies.

Breitenbach et al. in their research on the impact of Information and Communication Technologies on South Africa argue that these technologies have a significance impact on companies' productivity and that Information and Communication Technologies improve the mechanisms used by firms to retrieve, collect, transfer and manage huge amounts of data. This produces a direct positive outcome in the sense that it reduces costs related to processes in accordance with information management within the concerned companies (Breitenbach et al., 2005).The evolving software industry constitutes a path to a wide range of countries such as South Africa to enter the global Information Technology production market (Heeks,1996).

Heeks argues that figures tend to sustain that the use of software industry, more specifically exporting software products has contributed to the success of establishing strong growing industries in certain countries and hence this constitutes an evitable boost for their economic development. Among these countries, India can be picked as an example with statistics showing that in the past decade, precisely between the periods from 94 to 95, exporting software products generated 40 millions US dollars and this result grows by 4 percent or even more every year (Heeks, 1996).

From all these studies, it is undoubtedly clear that the software industry has been targeted by wide range of countries, especially new emerging industrial and developing countries mainly for its capacity to engender revenue either by means of exportation or any other value-added indicator that it constitutes for other industries. In the same perspective, it essentially requires efficient strategies and strong milestones for the effectiveness of the government in order to take local companies and firms in all other industries at a competitive dimension (Shirley et al., 2003).

Even for the most industrialized countries, software engineering is a potential factor for driving significant growth in economy; such powerful countries include for instance the United States of America and Microsoft is a tangible example of powerful resource that shows the impact of the software industry in this country's economic sector (Momodu et al., 2007).

The value of the information and communication technologies has been recognized as a significant component for economic upliftment by different administrative entities, ranging from government structures to public sector. In August 2008, while addressing delegates at the GovTech, South African Minister for Pubic Service and Administration, Geraldine Fraser Moleketi makes this statement: "Any incremental improvement in public services through ICT spends should positively impact on millions of people. Can we really say that is the case? How many ICT projects that have been delivered in the past few years have contributed positively to the millions of people in South Africa?"(ITWeb Informatica, 2009).

Pointing out the vision of the South African government on economy, Isaac Mophatlane, the Head of Public Sector at Business Connexion enthuses that "The South African Government is striving to be more citizen-centered and technology is the cornerstone that will deliver this vision"; he latter adds " Factors such as cost reduction, improved service delivery and citizen convenience are legitimate expectations Government intends to realize through investments in technology" (ITWeb Informatica, 2009).

There are growing numerous studies on the impact of the latest technologies on the development of the economic sectors of countries and thus, these researches tend to raise concerns about how to practically measure these technologies in terms of their direct contribution to sustaining economic development. In South Africa, the SAITIS project has set a framework of major indicators throughout the process of measuring how to achieve economic growth; among others, these indicators include the revenues to generated by Information Technology, the employment opportunities that will be created by this sector, their contribution to the Gross Domestic Product (GDP), the revenues and outcomes of exported products in the sector, the investment in terms of percentage in the field and the global Information Technology and software industry in the economy of the country.

South Africa, as an emergent democratic nation and growing economic force in Africa still have a log way to go and in the focus of this paper, we examine how the Information and Communication technologies and more specifically the software industry contribute to the economic development of the country by taking a cue form example of a successful country's framework in occurrence Ireland. This paper takes a closer look at the framework adopted by the latter country to boot its economic force through software industry and measure the possibility of applying it to South Africa.

### III. Research Framework (Methodology)

As previously stated, the main objective of this paper is to focus and closely look at the South African Industry and its implication on the country's economic development. In order to clearly accomplish this, a significant analysis of factors that need to be considered or have already been considered to contribute to the emergence of economic development is essential.

In order to productively tackle this, the framework used by Heavin et al. in investigating the impact of software industry on Irish economic growth will be adopted. That is, it comprises two main parts: the external or exogenous factors that impact on the process of building the software industry and endogenous or internal factors that contribute to the emergence success of the industry.

For the purpose of providing this paper with significant resources for its core objective, a number of other frameworks were consulted among others include Heeks (1996) which provides significant insights on procedures to follow in order to implement software industries in Africa, Hanna et al.(2003) who focus on National software Industry development…but as abovementioned, our main inspiring framework is the one used by Heavin et al. for their study of this industry in Ireland.

### III.1 Exogenous or External Factors

We refer to a certain category of factors as external or exogenous for their indirect influence on development of the Information and Communication Technology sector and these exogenous factors are hard to control. These exogenous factors comprise among the others characteristics such as the population of the country, the availability of natural resources, location…and characteristics linked to cultural attributes such as the language, the consideration of education and literacy…

Following official American government's reports available at http://www.statssa.gov.za, the South African population was estimated in 2007 being of 47.9 million and is composed of black in majority, white, colored and Asian. This factor in terms of population gives the country a predominance in industrial development as it provides a good workforce. The country presents a great range of natural resources that are favorable for its economic development, these resources include almost all necessary resources such as diamond, gold…and South Africa is actually the unique country is the world that produces fuel from coal. These resources enable an emergence of industries such as mining, machinery, motor vehicle manufacturing…

Due to its past experience with related crisis on the political aspect, the educational system of the country is still in transition since the government is still restructuring the all system to eliminate racial segregations and provide the same standards of education to all South Africans. This will offer a better environment for a proper emergence of a strong software industry.

Factors related to cultures are very important in participating to create a favorable atmosphere for an emergent software industry capable enough to enter into the international market. English, as an official language used in South Africa constitutes a major advantage to the development of software industry, not only that south Africans can trade internationally and gain access to a great range of countries world wide through the language, but also this has attracted big manufacturing and software development companies to set up in South Africa, to exemplify we will name Microsoft Corporation and other major international software providers such as Computer associates (CA)…A broader view of government's initiatives to improving the country's educational system during the post apartheid period, is a significant asset to the development and implementation of a market-competitive and active software industry.

**III.2 Endogenous or Internal Factors**

Endogenous factors refer to these attributes that can be created and adopted as milestones in developing Information and Communication Technology industry. The framework used to investigate the software industry in Ireland considers these factors at three different levels: national, enterprise and individual.

At national level, that is the implication of government in the industry. The regulations and policies that the latter formulate constitute a pipeline to supporting the development of the software industry. At the enterprise level, factors that facilitate the possibility of developing a successful software industry include for instance their vision and inspiration from other emerging enterprises, the willingness to take risks and develop strategic operations for their implication in the software market. At the individual level, essential factors that vitally impact on the establishment of a successful software industry include attributes such as the attitude and determination of South African to work, their motivation, desire of technological innovation, and the atmosphere at the information society. As stated previously technology and innovation are considered as vital to the economic growth.

**IV. Analysis and Contextualization**

**IV.1 National-level factors**

At the national level, an active implication from the government is required to support and implement programs, policies and concrete governmental actions with the intent to contribute to the establishment of an active software industry. It is clear that every country or nation that wish to focus their programs to vitalize the software industry as a key component of their economic development won't take the same path, but a clear and an efficient vision on resources distribution plays a great role on the best route to adopt. These resources include among others availability of required infrastructures and human force. English, as quoted previously, as the English used in South Africa is a vital key to international market and opportunities.

Most of the countries that have evolved in this industry have been working on the resources at their disposal and found themselves ways to enter the global market. But when it comes to allocating resources or deciding on mobilizing resources to build and support the industry, the government's action is the expected and prominent. Heeks support that

Dr Hanna points out that in countries that lack enough infrastructures, and encounter problems with their resources such as the case of Ireland at some point, need special initiatives from the government in order to boost the industry (Hanna et al., 2003).

In South Africa such initiatives are being encouraged by the government despite its any drawbacks that may block the process. In 2007, the South African minister of trade and industry talked about the implementation of the National Industrial Policy. This actual policy takes into account the role of ICT into helping achieve the required economic development rate set for 2010. The minister continues by clarifying that the principal approach of this particular policy is arrow government's actions on industrial development. He pointed out four critical areas that constitute the priority of this industrial policy: these include creating a favorable regulatory environment for macro-economy, enhance and mobilize necessary infrastructures, producing skilled and educated industrialists and bring effort for innovation in technology (dti minister, 2007).

As mentioned in the introduction the information technology field ,as sustained by a great number of published research reports , has constantly attracted the attention of the South Africa's government regarding its approach its potential contribution to economic growth and development(Breitenbach et al., 2005).

Another essential factor that supports and shows the involvement of the South African government is the SAITIS project. This project was developed to face the challenges of technologies in the economic field and create empowerment for all South Africans. In details the project has set a framework of major indicators throughout the process of measuring how to achieve economic growth; among others, these indicators include the revenues to generated by Information Technology, the employment opportunities that will be created by this sector, their contribution to the Gross Domestic Product (GDP), the revenues and outcomes of exported products in the sector, the investment in terms of percentage in the field and the global Information Technology and software industry in the economy of the country (Breitenbach et al., 2005).

In terms of funding and financial contribution or provisions of the South African government for the software industry, we will look at the department of trade and industry in order to get insights about how this is accomplished. In 2007, at the same occasion the minister talked about the industrial policy for enforcing development of software industry, he also pointed out that the software industry is recognized by the department of trade and industry as a vital component of the Information Technology sector and will become a valuable driving factor for enabling economic growth. He provided figures of the South African Software industry in terms of finances for the year 2006 that were estimated at 13 billion rand and this value was expected to increase over the next years. In terms of what this industry has brought to the South African economy just for its exported products, an amount that is valued being over R500 million is advanced, even tough it is less than the domestic market.

He quotes: "*The software industry is strongly pyramidal in structure regarding revenue concentration, with the five largest software vendors controlling approximately 40% of the software revenue and the ten largest controlling 50% of the revenue*". (Trade and Industry Minister, 2007).

In order to support the industry, education should be prioritized by the government and be considered as a key milestone to establishing competent personnel capable of operating in the software industry. Each year, when the South African financial minister Trevor A Manuel presents the budget at the parliament, education has always be considered among the vital sectors where considerable amounts of money are allocated. In his 2008 budget speech before the national assembly, the financial minister noted that an improved considerable amount of R46 billion was allocated to provinces for sectors such as health and education among others.

Indisputably, education is the critical path if not the most accurate and appropriate way to equip the society with required skills for any sector of economic development. In the same sense, education of women should be encouraged and even practical initiatives should be created to support and give opportunities to women especially in technical fields such as IT and Engineering. Such initiatives in South Africa are visible; an example to consider in this case is the Women in IT project that provides financial support to young South African ladies to pursue their studies in IT.

### IV.2 Enterprise-level factors

The second set of endogenous factors is associated with the enterprise. In the previous lines, we mentioned that the spirit that drives enterprises to create and promote innovative activities is a very important factor that sustains the emergence of software industry. With all these estimates of revenues produced in domestic software market of South Africa, we can understand how risk-taking software companies are in South Africa.

The South African Trade and Industry minister pints out that the South African Information and Communication Technology industry has made a significant impact within international perspective with its specialties in areas including "*electronic banking and financial services applications, fraud prevention systems and pre-payment*"(Trade and Industry Minister, 2007).

In this perspective, the South African government also engages in actions to support software companies in order to uplift their standards and capacitate them to enter global market competitiveness.

Opportunities offered by south African companies to young south African students that pursue their studies in the software development field particularly and even the Information Technology sector is vital factor to express the determination of these companies to have skilled personnel to support their software production.

### IV.3 Individual-level factors

Individual or human resource is the most critical asset to drive the development of a strong software development. The industry requires a team of skilled personnel trained to fulfill the wide sector of software development. The roles include for example programmers, software testers, software analysts, project managers…all these people constitute a vital milestone for the establishment of a successful software industry in South Africa.

Opportunities that are offered by software development companies in South Africa play a significant role in recruiting qualified people. The aim is to recruit a human resource equipped with strong entrepreneurial spirit, driven by a spirit of ingenuity and innovation and thus, forming the fundamental team of software development companies.

### V. Conclusion

Software industry as integral component of the Information and Communication Technology sector is vital to economic growth. Considering this industry to boost the economic sector of the Republic of South Africa requires a meticulous control and implementation of a set of factors. These include the intervention of the government in allocating enough resources for the implementation of the industry. Policies and regulations made in consideration of the pyramidal role that plays the South African Software Industry to economic development of the country. Enterprises and human resources are vital assets to establishing a successful software industry capable of entering the international market. And for all these to be concrete, the government needs wok in conjunction with industrial firms to promote initiatives that are intended to give opportunities for education, employment and finance these mechanisms to produce a workforce that is software-literate.